\documentclass[aps,prb,reprint,superscriptaddress]{revtex4-2}
\usepackage{csquotes}
\usepackage{amsmath}
\usepackage{amsfonts}
\usepackage{amssymb}
\usepackage{graphicx}
\usepackage{ragged2e}
\usepackage{subcaption}
\begin{document}

%\preprint{}

\title{Plexciton-mediated Raman scattering in strongly coupled systems}


\author{Danyang Liu}
\thanks{These authors contributed equally to this work.}
\affiliation{Institute of Light Resources and Environmental Science, Henan Academy of Sciences, Zhengzhou 450046, China}
\author{Meijuan Sun}
\thanks{These authors contributed equally to this work.}
\affiliation{School of Physics, Xi’an Jiaotong University, Xi’an 710049, China}
\author{Li Chen}
\email{Contact author: lichen@hnas.ac.cn}
\affiliation{Institute of Light Resources and Environmental Science, Henan Academy of Sciences, Zhengzhou 450046, China}
\affiliation{School of Physics, Xi’an Jiaotong University, Xi’an 710049, China}
\author{Hao Shen}
\email{Contact author: haoshen@hnas.ac.cn}
\affiliation{Institute of Light Resources and Environmental Science, Henan Academy of Sciences, Zhengzhou 450046, China}
\affiliation{School of Physics, Xi’an Jiaotong University, Xi’an 710049, China}

\date{\today}

\begin{abstract}
A series of experimental results demonstrate a distinctive Raman response in plasmon-exciton coupled systems. The enhancement of Raman scattering varies for different phonon modes.
We describe the microscopic dynamical process of this Raman scattering using quantum many-body theory.
Unlike conventional Raman scattering involving electron-phonon interactions, the process in plasmon-exciton coupled systems is characterized by inelastic scattering between phonons and plasmon-exciton polaritons---formed through the coupling of plasmons and excitons---acting as intermediate states. We derive analytical expressions for the Raman intensity and enhancement factors for various phonon modes, which show excellent agreement with experimental data.
Furthermore, experimental fittings indicate a substantial disparity in the linewidths of the upper and lower polariton branches, for which we provide a comprehensive theoretical explanation.
Based on linear response theory, we propose a microscopic mechanism for the formation of plasmon-exciton polaritons, enabling the analytical calculation of their dispersions and linewidths. This approach naturally accounts for the significantly asymmetry observed in the linewidths of the upper and lower polariton branches. By characterizing the polariton-phonon scattering process at the quantum level, we reveal the fundamental physical mechanism driving polariton-enhanced Raman scattering. Our work establishes a universal framework for describing the dynamical evolution of coupled systems, providing a versatile paradigm for exploring the interactions between plasmons and other quasiparticles.
\end{abstract}

%\keywords{}

\maketitle

\section{introduction}
Plasmons are the quantized modes of collective electronic oscillations, a concept first proposed by Pines et al.~\cite{bohm1951collective,pines1952collective,bohm1953collective,pines1956collective,pines2018elementary}.
Pines and his colleagues discovered that when a charge perturbation exists within an electron gas, the remaining electrons move to screen the perturbation and reduce the net electric field due to the Coulomb interactions between them. For electron interactions at scales smaller than the Debye length, true screening occurs, however, interactions at scales larger than the Debye length lead to the collective oscillation of electrons. These large-scale collective oscillations are defined as plasmons~\cite{pines1956collective}.
To describe these fluctuations in charge density, Pines defined a quantum mechanical operator for charge density. By calculating the commutator between the charge density operator and the system's Hamiltonian, he derived the equations of motion for the operator, thereby determining the characteristic oscillation frequency of the electrons' collective motion.
The calculation results indicate that the oscillation frequency of plasmons is directly related to the electron density of the system. For typical solids, the plasmon energy is on the order of several to tens of electron volts. Since this energy scale is significantly greater than that provided by thermal motion, the internal thermal motion of the metal cannot induce this collective oscillation mode, and exciting plasmons requires external energy~\cite{pines1956collective}.

Plasmons were originally proposed to explain the energy loss of electrons passing through thin solid films, therefore, electron energy loss spectroscopy (EELS) can serve as an indirect method for observing plasmons~\cite{eberlein2008plasmon,liu2008plasmon,lu2009plasmon,roman2017low,wachsmuth2013high}. This technique reveals the energy and dispersion relation of plasmons in momentum space by analyzing the energy loss of inelastically scattered electrons.
In recent years, with breakthroughs in nano-optical measurement techniques, particularly the development of near-field optical microscopy, real-space imaging of plasmons has been achieved~\cite{hillenbrand2002material,chen2012optical,fei2012gate}, allowing direct observation of plasmon propagation and interference on material surfaces. Most notably, the wavelength of excited plasmons in materials is significantly shorter than that of the incident light. This extremely short wavelength signifies the ability of plasmons to confine optical fields to subwavelength scales. Leveraging this field confinement property, plasmonic cavities with highly compressed mode volumes can be constructed, and their resonant characteristics can be tuned in situ by applying an external gate voltage~\cite{schuller2010plasmonics,chikkaraddy2016single,baumberg2019extreme,elliott2022fingerprinting}.
While the experimental studies mentioned above have provided compelling evidence for the existence of plasmons, from momentum space to real space, these measurements themselves have not revealed the underlying microscopic physical mechanisms. The microscopic excitation mechanism of the collective motion of numerous electrons in interacting electron systems has not been fully described. Furthermore, the intrinsic dispersion relation of plasmons and their characteristic width, as well as the resulting optical field confinement and enhancement effects, have not been thoroughly discussed theoretically. To describe the excitation and propagation of plasmons at the microscopic level, we need to introduce the framework of linear response theory. This theory establishes the constitutive relationship between an external perturbation and the resulting electron density response of the system, thereby rigorously defining the physical picture of plasmons as collective excitation modes of charge density fluctuations~\cite{pines1966theory,giuliani2008quantum,mahan2013many}.

The remarkable light-confinement capability of plasmons enables the concentration of electromagnetic energy into nanoscale volumes far below the excitation wavelength-specifically within the \enquote{hot spots} of nanogaps~\cite{aizpurua2005optical,marinica2012quantum,wang2020fundamental,lu2024quantifying}. This resulting giant local field enhancement provides a robust framework for manipulating light at the nanoscale and significantly amplifies light-matter interactions. However, constrained by intrinsic ohmic losses in metals, plasmons exhibit extremely short lifetimes, posing challenges for long-range energy transport and long-term storage. Consequently, constructing hybrid systems that couple plasmons with other quasiparticles---notably exciton-plasmon coupled systems---has emerged as a prominent research focus. Excitons are bound states of excited electrons and holes held together by Coulomb interactions; in transition metal dichalcogenides (TMDCs), strong confinement in the thickness dimension leads to binding energies reaching hundreds of millielectronvolts, allowing excitons to remain stable at room temperature~\cite{chernikov2014exciton,he2014tightly,ye2014probing,wang2018colloquium,mueller2018exciton}. By leveraging complementary strengths, the plasmon-exciton system integrates the velocity of photons, the precision of nanoscale localization, and the long coherence times of excitons, offering an unprecedented platform for tailoring light-matter interactions~\cite{kleemann2017strong,stuhrenberg2018strong,schneider2018two}. Elucidating the underlying microscopic physical mechanisms of this coupling is therefore of paramount importance.
In previous studies of plasmon-exciton interactions, it has been common practice to simplify both plasmons and excitons as distinct quasiparticles, employing an effective Hamiltonian to phenomenologically describe their coupling~\cite{hopfield1958theory,torma2015strong,hwang2015quantum}. However, this approach fails to elucidate the microscopic dynamical mechanisms underlying plasmon generation or the fundamental nature of their interaction. Furthermore, effective models struggle to accurately yield the dispersion relations and intrinsic linewidths of plasmons, nor can they precisely characterize the dynamical evolution of polaritons within the coupled system. This inherently limits the accuracy of quantitative analyses of experimental data. In contrast, a theoretical framework grounded in linear response theory allows for a microscopic clarification of plasmon excitation mechanisms. By evaluating the total response function of the coupled system, one can rigorously derive the polariton dispersion and damping rates. This many-body approach to describing quasiparticle interactions has been successfully implemented in the study of plasmon-phonon coupling systems.

Raman scattering is an inelastic light-matter interaction process whose spectral signatures record the energy exchange between incident photons and the quasiparticle system~\cite{raman1928new,guntherodt1975light}. 
In plasmon-exciton coupled systems, the hybridized polariton states not only significantly enhance the Raman scattering cross-section through strong local field amplification, but also serve as intermediate states for phonon scattering, fundamentally restructuring the system's energy-level structure and Raman response characteristics.
We provide a quantum-mechanical description of the microscopic polariton Raman scattering process, deriving a detailed formula for the Raman intensity involving polariton-mediated phonon scattering and elucidating the evolution of Raman signals across various coupling regimes. In contrast to conventional Raman theories~\cite{henry1965raman,burstein1968raman,bendow1970polariton,as1972response,zeyher1974polariton,ushioda1983raman}, we present an analytic expression for the polariton Raman enhancement factor, which demonstrates excellent agreement with experimental data. Furthermore, our experimental fits reveal a pronounced asymmetry in the linewidths of the upper and lower polariton branches. Through precise calculations, we account for the drastic disparity in the damping rates between these branches, thereby uncovering the physical origin of the distinct decay dynamics inherent to different polariton states.

In this work, we elucidate the microscopic mechanism underlying the generation of collective electronic oscillations within interacting electron systems, providing a rigorous derivation of the plasmon dispersion relation. In a coupled system, modulating the dimensions of metallic nanostructures essentially tunes the characteristic resonant wave vectors between plasmons and other quasiparticles. By identifying the poles of the total response function for the plasmon-exciton coupled system, we precisely determine the polariton dispersion curves. Furthermore, by accounting for the imaginary part of the response function, we derive the intrinsic linewidths and damping characteristics of the polaritons. These theoretical derivations demonstrate excellent agreement with experimental observations, validating our microscopic framework.

\section{plasmon dispersion}
The generation of plasmons is the response of an interacting electron gas to an external field. This collective behavior can be described by linear response theory. Response refers to the deviation of certain observable physical quantities from their equilibrium state after the system is perturbed. The response function is the core of linear theory, it not only establishes a quantitative relationship between the physical quantities of interest and the external perturbation but also reveals the dynamic charateristics of the system's response process. In the study of plasmons. the charge density $n(\vec{r})$ is typically defined to describe the charge distribution and its dynamic fluctuations~\cite{pines1956collective,giuliani2008quantum}. The charge density is linked to the external perturbation through the density-density response function. As plasmons are the charge density oscillation modes of the electron system, the analytical structure of the density-density response function determines the existence and dynamical behavior of the plasmons. While external perturbations can excite plasmons, the dispersion charateristics of plasmons depend solely on the intrinsic properties of the material and are independent of the form of the external perturbation.

Exciting plasmons with an optical field is the most common method. When an external light field illuminates a metal surface. the electron system in the metal responds to the perturbative potential. resulting in charge density fluctuations. According to linear reponse theory, the induced charge density is linearly dependent on the external perturbation, with the density-density response function mediating this relationship. The Coulomb potential generated by the induced charge superimposes on the external potential to form the total potential of the system, known as the screened scalar potential. The screened scalar potential differs from the external perturbative potential by the dielectric function $\epsilon(q,\omega)$,
\begin{equation}
\epsilon(q,\omega)=1-v_{\mathrm{c}}\tilde{\chi}_{nn}(q,\omega),
\end{equation}
where $v_{\mathrm{c}}=2\pi e^{2}/q$ is the Coulomb potential in two-dimensional momentum space, and $\tilde{\chi}_{nn}(q,\omega)$ is the proper density-density response function~\cite{giuliani2008quantum,das2009collective}. Specifically, the screened scalar potential is the external perturbative potential divided by the dielectric function. Therefore, the zeros of the dielectric function represent a massive enhancement of the system's overall potential energy, signifying a maximal response of a large number of electrons to the external perturbation. The zeros of the dielectric function---or equivalently, the poles of the proper density-density response function-yield the plamson dispersion relation~\cite{hwang2007dielectric,jablan2009plasmonics,hwang2010plasmon}. The proper density-density response function is determined by the electronic structure of the material.

Under the Random Phase Approximation (RPA), the proper density-density response function is usually approximated by the non-interacting density-density response function, namely the Lindhard function. In the long-wave limit $(q\rightarrow 0)$. the two-dimensional proper density-density response function is approximately expressed as~\cite{giuliani2008quantum}:
\begin{equation}
\tilde{\chi}_{nn}^{\mathrm{RPA}}(q,\omega)=\dfrac{nq^{2}}{m\omega^{2}}\left(1+a_{d}\dfrac{q^{2}v_{F}^{2}}{\omega^{2}}\right),
\end{equation}
where $n$ is the electron density of the metal nanoparticles, $m$ is the electron mass, $v_{F}$ is the Fermi velocity of the eletron in metal nanoparticles, which is directly related to the electron density $n$, and $a_{d}=3/4$ is a constant factor.

The plasmon dispersion relation is given by the zeros of the dieletric function. Solving the equation $\epsilon(q,\omega)=0$, we obtain the expression for the plasmon dispersion relation,
\begin{equation}
\omega_{\mathrm{pl}}
=\omega_{p}\left(1+a_{d}\dfrac{q}{\kappa_{2}}\right),
\label{eq:w_pl}
\end{equation}
where $\omega_{p}=\sqrt{\dfrac{2\pi ne^{2}q}{m}}$, $\kappa_{2}=\dfrac{4\pi ne^{2}}{mv_{F}^{2}}$.

Normalized plasmon dispersion relation can be expressed as $\tilde{\omega}_{\mathrm{pl}}=\tilde{\omega}_{F}\sqrt{\tilde{q}}\left(1+a_{d}\dfrac{\tilde{q}}{\tilde{\kappa}_{2}}\right)$, where we define the normalized variables $\tilde{\omega}_{\mathrm{pl}}=\dfrac{\hbar\omega_{\mathrm{pl}}}{\epsilon_{F}},\ \tilde{q}=\dfrac{q}{k_{F}}$ and constants $\tilde{\omega}_{F}=\dfrac{2e}{\hbar}\sqrt{\dfrac{m}{k_{F}}},\ \tilde{\kappa}_{2}=\dfrac{2me^{2}}{\hbar^{2}k_{F}}$.
\begin{figure}[h]
\centering
\includegraphics[width=0.45\textwidth]{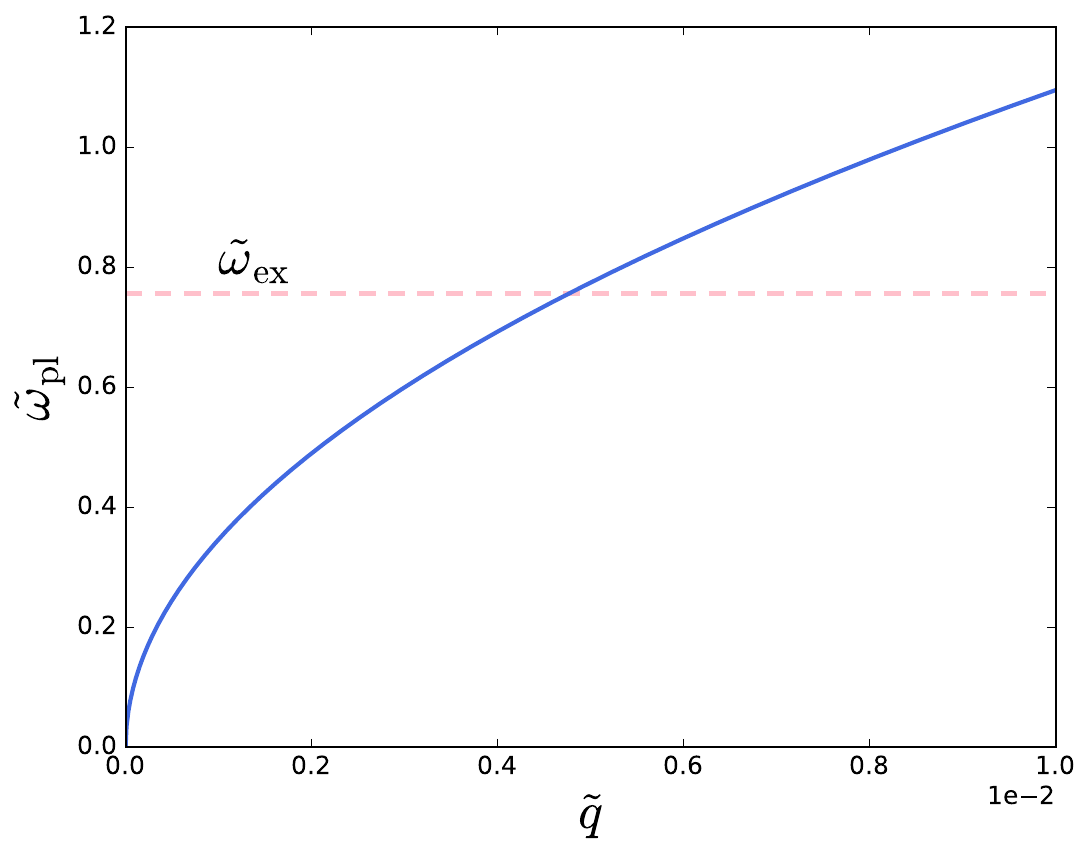}
\caption{\justifying Normalized plasmon dispersion relation.
In the study of strong coupling systems using nanoparticle-on-mirror (NPoM) microcavities~\cite{liu2021nonlinear}, the metallic nanoparticle supports the plasmons, while the two-dimensional material within the cavity provides the excitons. Experimentally, by precisely tailoring the geometric dimensions of the metallic nanoparticle, one can effectively tune the plasmonic resonance wavevector, thereby selecting a specific plasmonic mode to participate in the strong coupling. In contrast, the exciton energy of the material is typically fixed, taking monolayer molybdenum disulfide ($\mathrm{MoS_{2}}$) as an example, its exciton resonance energy is fixed at $1.81\ \mathrm{eV}$~\cite{park2018direct}, as indicated by the pink dashed line. When the energy of plasmon approaches the exciton energy, strong coupling occurs between the plasmons and excitons, leading to the formation of exciton-plasmon polaritons.}
\label{fig:disp_pl_nor}
\end{figure}
%密度-密度响应函数联系起诱导电荷密度与外部微扰势，而固有密度-密度响应函数联系起诱导电荷密度与屏蔽势。

\begin{figure}[h]
\centering
\begin{subfigure}[b]{0.45\textwidth}
\centering
\includegraphics[width=\textwidth]{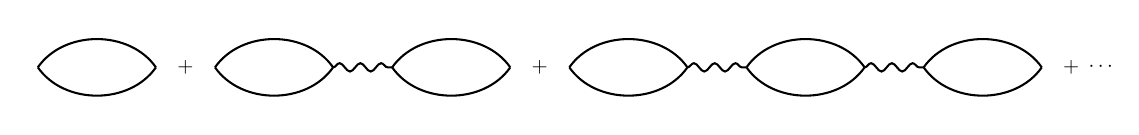}
\caption{Photon mediated response function}
\label{fig:chi_gamma}
\end{subfigure}
\hfill
\begin{subfigure}[b]{0.45\textwidth}
\centering
\includegraphics[width=\textwidth]{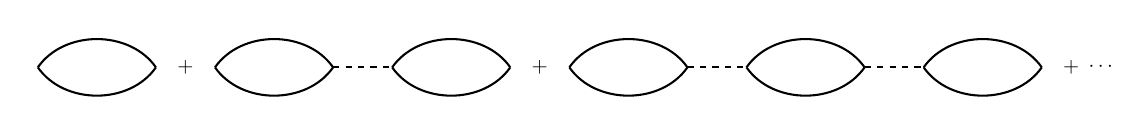}
\caption{Exciton mediated response function}
\label{fig:chi_exciton}
\end{subfigure}
\caption{\justifying Infinite series of Feynman diagrams corresponding to the RPA for density-density response function. In Figure (a) The leading-order term, represented by a closed bubble diagram, corresponds to the non-interacting response function. Subsequent terms consist of polarization bubbles connected by wavy lines representing the Coulomb potential. This self-consistent summation of infinite-order ring diagrams physically characterizes the collective excitations emerging from dynamical screening effects. In Figure (b), this framework is extended to a composite coupled system, where the introduction of exciton polarization modifies the bare Coulomb interaction to construct the total response function. The collective modes of the system are determined by the zeros of the total dielectric function, whose characteristic solutions reveal the polariton dispersion under strong plasmon-exciton coupling.}
\label{fig:chi}
\end{figure}

The density-density response function $\chi_{nn}(q,\omega)$ is, in fact, the statistical average of the density-density commutator over a thermal equilibrium ensemble. In perturbative calculations of linear response functions, the causual response function can be expressed as a time-ordered correlation function. These correlation functions of density operators can be represented by Feynman diagrams. Figure~\ref{fig:chi_gamma} illustrates the time-ordered correlation function of the density operators for an interacting electron system~\cite{hwang2010plasmon}. In Figure~\ref{fig:chi_gamma}, the bubble diagram in the first term represents the non-interacting correlation function $\tilde{\chi}_{nn}(q,\omega)$. This bubble diagram is formed by joining two directed solid lines, where each solid line represents a non-interacting Green function. The wavy lines in the subsequent diagrams represent the Coulomb potential $v_{\mathrm{c}}$ between electrons. Summing all the terms in Figure~\ref{fig:chi_gamma} yield the density-density response function corrected by first-order electron-electron interactions,
\begin{equation}
\chi_{nn}(q,\omega)=\dfrac{\tilde{\chi}_{nn}(q,\omega)}{1-v_{\mathrm{c}}\tilde{\chi}_{nn}(q,\omega)}=\dfrac{\tilde{\chi}_{nn}(q,\omega)}{\epsilon(q,\omega)}.
\label{eq:chi_nn_RPA}
\end{equation}
When calculating the response function using perturbative methods. the random phase approximation (RPA) only accounts for the response function corrected by the first-order Coulomb interaction between electrons~\cite{giuliani2008quantum}. The full response function and the proper response function differ by the dielectric function. The screened scalar potential is the ratio of the external perturbative potential to the dielectric function. As discussed previously, the zeros of the dielectric function reflect the system's maximal response to the perturbation.
Typically, electron-electron interactions refer to Coulomb interactions, where one electron experiences the Coulomb potential generated by another. In the dynamics of quantuam electric dynamics theory, the Coulomb force is essentially a process where eletrons interacts via the exchange of virtual photons. The wavy lines in the Feynman diagrams can be regarded as photon propagators, with the Coulomb potential serving as the scalar component of the Fourier-transformed photon propagators.

A plasmon is a type of collecitve oscillation, representing the coherent vibration of the electron gas as a whole relative to the ionic background. This mode can be self-sustaining even in the absence of external perturbations. Mathematiclly, the condition for a system to support such self-sustained oscillation is that an internal density fluctuation can be generated even when the external perturabtion is zero. According to linear response theory, the repsonse function is the key to linking the response to the external perturbative potential, specifically, the density response is the product of density-density response function and the external perturbation, expressed as $n_{\mathrm{ind}}=\chi_{nn}V_{\mathrm{ext}}$. Consequently, an intrinsic oscillation exists within the system only when the density response function diverges, which corresponds to the existence of poles. The reducible response function differs from the non-interacting response function by the dielectric function $\epsilon(q,\omega)$. Since $\tilde{\chi}_{nn}(q,\omega)$ is typically finite, the expression can diverge only when the denominator vansihes---that is, when the dielectric function equals zero.

\section{plasmon-exciton polariton dispersion}
As a quasiparticle, a plasmon can interact with other particles or quasiparticles. Plasmons are defined as the collective oscillations of electrons, which originate from the non-negligible Coulomn interactions between them. These Coulomb interactions result from the exchange of virtual photons between eletrons.
When considering the interaction between plasmons and excitons, we must account for not only the process of force transmission via virtual photon exchange but also the interaction mediated by the exchange of excitons. The exciton propagator corresponds to the scalar potential $v_{\mathrm{ex}}=g^{2}/(\omega-\omega_{\mathrm{ex}})$~\cite{antonius2022theory}, where $g$ represents the electron-exiton coupling constant and $\omega_{\mathrm{ex}}$ denotes the exciton energy.
The feynman diagram show in Figure~\ref{fig:chi_exciton} illustrates the response function contributed by the interaction between electron through exciton exchange. Together Figure~\ref{fig:chi_gamma} and Figure~\ref{fig:chi_exciton} provide the complete response function of the system,
\begin{equation}
\dfrac{\tilde{\chi}_{nn}(q,\omega)}{1-v_{\mathrm{c}}\tilde{\chi}_{nn}(q,\omega)}
+\dfrac{\tilde{\chi}_{nn}(q,\omega)}{1-v_{\mathrm{ex}}\tilde{\chi}_{nn}(q,\omega)}
=\dfrac{\tilde{\chi}_{nn}(q,\omega)}{\epsilon_{\mathrm{eff}}(q,\omega)},
\label{eq:epsilon_eff}
\end{equation}
where the effective dielectric function can be expressed as $\epsilon_{\mathrm{eff}}(q,\omega)=1/2-(v_{\mathrm{c}}+v_{\mathrm{ex}})\tilde{\chi}_{nn}(q,\omega)/2$.
The zeros of this effective dieletric function yield the dispersion relation of the plasmon-exciton polaritons. By solving the equation $\epsilon_{\mathrm{eff}}(q,\omega)=0$, we obtain:
\begin{equation}
\omega_{\mathrm{pol}^{\pm}}=\dfrac{3\omega_{\mathrm{ex}}}{4}+\dfrac{\omega_{\mathrm{pl}}^{2}}{4\omega_{\mathrm{ex}}}\pm\sqrt{\dfrac{(\omega_{\mathrm{ex}}^{2}-\omega_{\mathrm{pl}}^{2})^{2}}{16\omega_{\mathrm{ex}}^{2}}+\dfrac{ng^{2}q^{2}}{2m\omega_{\mathrm{ex}}}}.
%\tilde{\omega}_{pol\pm}=\dfrac{3\tilde{\omega}_{\mathrm{ex}}}{4}+\dfrac{\tilde{\omega}_{\mathrm{pl}}^{2}}{4\tilde{\omega}_{\mathrm{ex}}}\pm\sqrt{\dfrac{(\tilde{\omega}_{\mathrm{ex}}^{2}-\tilde{\omega}_{\mathrm{pl}}^{2})^{2}}{16\tilde{\omega}_{\mathrm{ex}}^{2}}+\dfrac{\hbar ng^{2}\tilde{q}^{2}}{2m\epsilon_{F}^{2}\tilde{\omega}_{\mathrm{ex}}}}.
\end{equation}

Plasmons are collective oscillations of interacting electrons, driven primarily by non-negligible inter-electronic Coulomb forces. When we account for interactions mediated not only by the exchange of virtual photons, the Coulomb force but also by the exchange of excitons between electrons, the resulting collective oscillation can be viewed as an excitation formed by the coupling of plasmons and excitons. This excitation can be defined as a new quasiparticle: the plasmon-exciton polariton or plexciton~\cite{perez2013optical}. The formula gives the dispersion curves of the upper and lower polariton branches. This antisymmetric dispersion relation has been reported in many literature~\cite{huang1951interaction,hopfield1958theory,weisbuch1992observation,liu2015strong,torma2015strong,chikkaraddy2016single}.

\begin{figure}[h]
\centering
\includegraphics[width=0.45\textwidth]{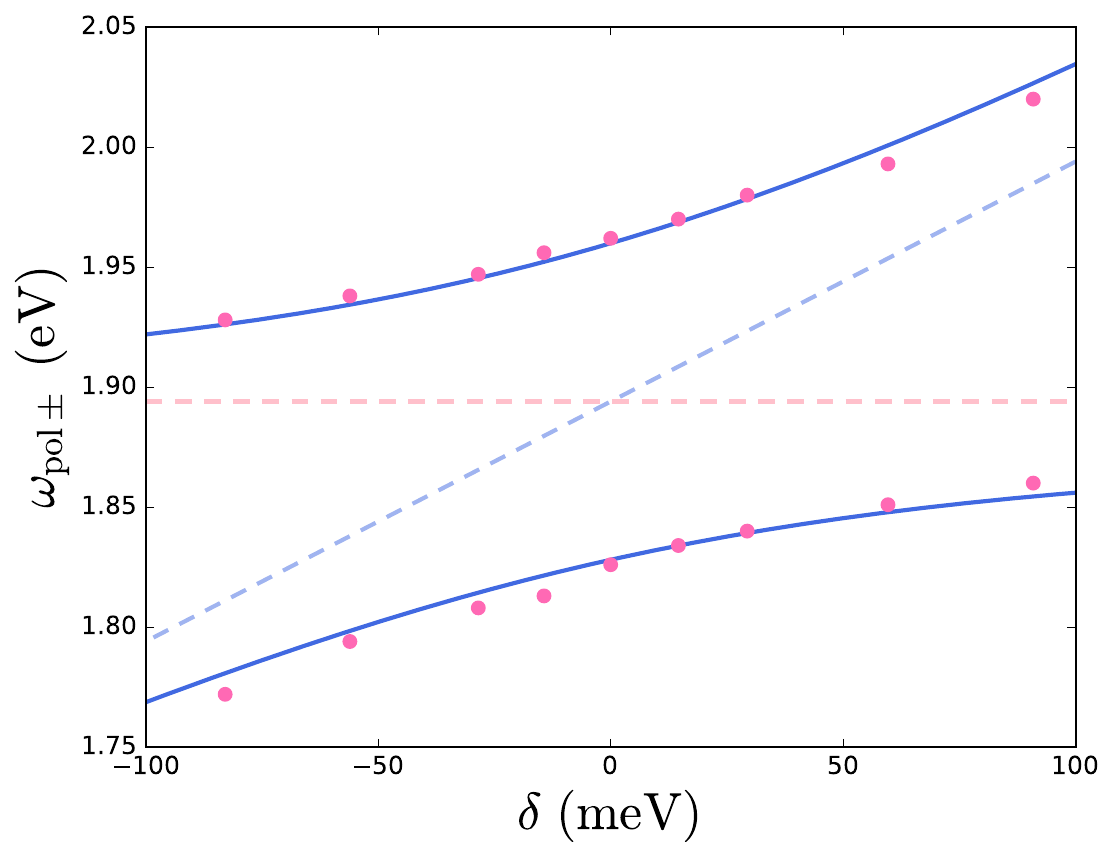}
\caption{\justifying Plasmon-exciton polariton dispersion relation. The variation of the upper and lower polariton branch energies as a function of detuning. The blue solid lines represent the two polariton branches, the pink dashed line indicates the exciton energy, and the blue dashed line shows the plasmon dispersion. The red dots denote our simulation results. The polariton branches exhibit clear anticrossing characteristics, reflecting the strong coupling between the exciton and the plasmon.}
\label{fig:disp_pol_nor}
\end{figure}

Typically, the plasmon that strongly couples with the exciton has a relatively small resonant wavevector. Experimentally, researchers are more concerned with the variation of the polariton branches as a function of detuning. We define the energy difference between the plasmon and the exciton as the detuning, $\delta = \omega_{\mathrm{pl}} - \omega_{\mathrm{ex}}$. When the detuning is zero ($\delta = 0$), the sum of the energies of the upper and lower polariton branches equals exactly twice the exciton energy, i.e., $\omega_{\mathrm{pol+}} + \omega_{\mathrm{pol-}} = 2\omega_{\mathrm{ex}}$.
Figure~\ref{fig:disp_pol_nor} shows the dispersion relation of the upper and lower polariton branches as a function of detuning. In a given experimental system, the exciton energy is fixed; therefore, the detuning is entirely determined by the plasmon energy. The energy of the plasmon mode, in turn, is determined by its resonant wavevector. Consequently, the dispersion relation of the upper and lower polariton branches with respect to detuning essentially reflects the variation of these two branch energies with the resonant wavevector, also exhibiting an anti-crossing characteristic.

The plasmon-exciton polariton is a composite particle formed by the coupling of plasmon and exciton. This composite excitation combines the characteristics of both plasmons and excitons, leading to novel light-matter interaction phenomena. Such composite excitations can be treated as quasiparticles that scatter with phonons in the material; their unique Raman signatures reflect the specific coupling modes between plasmons and excitons.

\section{plexciton raman scattering}
Raman scattering is a process in which light undergoes inelastic scattering with matter, thereby altering the photon frequency. When the energy of the incident photon is significantly lower than the bandgap energy of the material, electrons can transition to virtual states. Numerous virtual eletronic states participate collectively as intermediates states in the scattering process, rending the specific details of each individual state inconsequential. In this case, the spectrum depends on the symmetry of the phonon modes, such a process is termed convential Raman or non-resonant Raman scattering. Non-resonant Raman scasttering only provides information regrading phonons and rarely reflects details such as electronic structure or eletron-phonon coupling.
When the energy of either the incident or scattered light precisely matches a specific energy level in the electronic structure or an excitionic transition level of the material, the Raman process becomes a scattering event dominated by a few intermediated states. This is known as resonant Raman scattering. In this regime, the Raman signal is significantly enhanced, and the selection rules originally determined by phonon symmetry may no longer hold. Resonant Raman signals are strongly dependent on the energy level structure of the system, providing not only phonon information but also reflecting details of the electronic levels and electron-phonon coupling. By the same token, when the incident photon energy approaches a polariton eneryg level, polariton resonance Raman scattering occurs. With polaritons serving as intermediate states, their hybridized nature signifiicantly boosts the Raman efficiency.

\begin{figure}[h]
\centering
\includegraphics[width=0.4\textwidth]{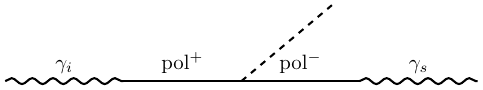}
\caption{\justifying Feynman diagram for polariton Raman scattering. The incident photon energy resonates with the polariton energy level, causing the system to transition into a real polariton state. Subsequently, this polariton state scatters with a phonon, transferring part of its energy to the phonon, and the system transitions further to the lower branch of the polariton states. Finally, the reduced-energy polariton emits a scattered photon, and the system returns to the ground state. The corresponding Feynman diagram contains two types of interaction vertices: the photon-polariton interaction vertex and the polariton-phonon interaction vertex.}
\label{fig:RamanFD}
\end{figure}

The occurence of conventional Raman scattering primarily involves three microscopic processes: the annihilation of an incident photon, the scattering of the phonon, and the generation of the scattered photon. The intermediate states in conventional Raman scattering are virtual electronic states. Through eletron-phonon coupling, these virtual states transfer part of their energy to the vibrational modes.
In contrast, the intermediate states in polariton Raman scattering are real plasmon-exciton polariton states. The microscopic process of polariton resonant Raman scattering can be divieded into the following key steps:1.Initial state:The initial state of the system consists of the electronic system in its ground state and the incident photon, $|\Psi_{i}\rangle=|g\rangle\otimes|\hbar\omega_{\gamma_{i}}\rangle$, $E_{i}=E_{g}+\hbar\omega_{\gamma_{i}}$. 2.Excitation of plasmons and excitons: When the incident light illuminates the coupled plasmon-exciton system, a plasmonic cavity, electrons in the metal undergo collective oscillations driven by the optical electric field to form plasmons. Simultaneously, electrons in the semiconductor are excited to the conduction band, leaving holes in the valence band to form electron-hole pairs, known as excitons. When the incident photon energy is close to the energies of both the plasmons and excitons, the system enters the strong coupling regime, where plasmons and excitons strongly hybridize to form plasmon-exciton polaritons. The polariton dispersion exhibits an anti-crossing behavior, resulting in upper and lower branches. The polariton serves as the intermediate state $|\Psi_{n}\rangle=|\hbar\omega_{\mathrm{pol}+}\rangle$; here, we assume the system is in the upper polariton branch, such that $E_{n}=\hbar\omega_{\mathrm{pol}+}$. 3.Polariton-Phonon Coupling: The polariton couples with a phonon, transferring part of its energy to the lattice. The polariton itself undergoes a transition from a higher energy state to a lower energy state, yielding $|\Psi_{n'}\rangle=|\hbar\omega_{\mathrm{pol}-}\rangle\otimes|\hbar\omega_{\mathrm{phonon}}\rangle$, $E_{n'}=\hbar\omega_{\mathrm{pol}-}+\hbar\omega_{\mathrm{phonon}}$. 4.Radiative Decay: The polariton state at the lower energy level relaxes via radiative decay to release a scattered photon. The final state of the system is $|\Psi_{f}\rangle=|g\rangle\otimes|\hbar\omega_{\gamma_{s}}\rangle\otimes|\hbar\omega_{\mathrm{phonon}}\rangle$, $E_{f}=E_{g}+\hbar\omega_{\gamma_{s}}+\hbar\omega_{\mathrm{phonon}}$. The polariton Raman scattering amplitude can then be expressed as,
\begin{equation}
\mathcal{M}=\frac{\langle\Psi_{f}|H_{\mathrm{pol}-\gamma}|\Psi_{n'}\rangle\langle\Psi_{n'}|H_{\mathrm{pol-ph}}|\Psi_{n}\rangle\langle\Psi_{n}|H_{\mathrm{pol}-\gamma}|\Psi_{i}\rangle}{(E_{i}-E_{n}+i\Gamma_{\mathrm{pol}+})(E_{f}-E_{n'}+i\Gamma_{\mathrm{pol}-})}.
\end{equation}
We now set the energy of the ground-state electrons $E_{g}$ to zero and assume that the incident photon, the scattered photon, and the phonon satisfy energy conservation, $\hbar\omega_{\gamma_{i}}=\hbar\omega_{\gamma_{s}}+\hbar\omega_{\mathrm{phonon}}$, thus, $E_{i}-E_{n}=E_{g}+\hbar\omega_{\gamma_{i}}-\hbar\omega_{\mathrm{pol}+}=\hbar\omega_{\gamma_{i}}-\hbar\omega_{\mathrm{pol}+}$,
$E_{f}-E_{n'}=E_{g}+\hbar\omega_{\gamma_{s}}+\hbar\omega_{\mathrm{phonon}}-\hbar\omega_{\mathrm{pol}-}-\hbar\omega_{\mathrm{phonon}}=\hbar\omega_{\gamma_{s}}-\hbar\omega_{\mathrm{pol}-}=\hbar\omega_{\gamma_{i}}-\hbar\omega_{\mathrm{phonon}}-\hbar\omega_{\mathrm{pol}-}$.
At the same time, we represent the transition matrix elements as the coefficients $g_{\mathrm{pol}-\gamma}=\langle\Psi_{f}|H_{\mathrm{pol}-\gamma}|\Psi_{n'}\rangle$, $g_{\mathrm{pol-ph}}=\langle\Psi_{n'}|H_{\mathrm{pol-ph}}|\Psi_{n}\rangle$, $g_{\mathrm{pol}-\gamma}=\langle\Psi_{n}|H_{\mathrm{pol}-\gamma}|\Psi_{i}\rangle$, here, $H_{\mathrm{pol}-\gamma}$ and $H_{\mathrm{pol}-ph}$ denote the plexciton-photon and plexciton-phonon interaction Hamiltonians, respectively. In this simplified analysis, the polarization dependence of the Raman scattering intensity is neglected, allowing us to isolate solely the effective coupling coefficients that characterize the quasiparticle interaction strength~\cite{pino2015quantum,gontijo2019double}. Then, Raman amplitude can be expressed as,
\begin{equation}
\mathcal{M}=\frac{g_{\mathrm{pol}-\gamma}g_{\mathrm{pol-ph}}g_{\mathrm{pol}-\gamma}}{(\hbar\omega_{\mathrm{pol}+}-\hbar\omega_{\gamma_{i}}-i\Gamma_{\mathrm{pol}+})(\hbar\omega_{\mathrm{pol}-}-\hbar\omega_{\gamma_{s}}-i\Gamma_{\mathrm{pol}-})}.
\end{equation}
The Raman intensity is proportional to the square of the Raman scattering amplitude,
\begin{equation}
I=\frac{g_{\mathrm{pol}-\gamma}^{2}g_{\mathrm{pol-ph}}^{2}g_{\mathrm{pol}-\gamma}^{2}}{[(\hbar\omega_{\mathrm{pol}+}-\hbar\omega_{\gamma_{i}})^{2}+\Gamma_{\mathrm{pol}+}^{2}][(\hbar\omega_{\mathrm{pol}-}-\hbar\omega_{\gamma_{s}})^{2}+\Gamma_{\mathrm{pol}-}^{2}]}.
\end{equation}

When the detuning $\delta=0$, the polariton branch energies are $\omega_{\mathrm{pol}+}=1.962\ \mathrm{eV}$ and $\omega_{\mathrm{pol}-}=1.826\ \mathrm{eV}$. We choose the incident light to be in resonance with the upper polariton branch, i.e., $\omega_{\gamma_{i}}=\omega_{\mathrm{pol}+}$. For a specific phonon mode, energy conservation dictates that $\omega_{\gamma_{s}}+\omega_{\mathrm{ph}}=\omega_{\gamma_{i}}$. In the expression, only the interaction coupling constants between quasiparticles, $g_{\mathrm{pol}-\gamma},\ g_{\mathrm{pol-ph}},\ g_{\mathrm{pol}-\gamma}$ and polariton width $\Gamma_{\mathrm{pol}+},\ \Gamma_{\mathrm{pol}-}$ remain as unknow parameters.

The Raman scattering intensity of polaritons is analogous to that of conventional Raman scattering; however, the intermediate state in the conventional Raman process is an electron rather than a polariton~\cite{gontijo2019double}.
Consequently, the Raman enhancement factor is defined as the ratio of the resonance Raman scattering intensity of the polariton to the ordinary Raman scattering intensity.
\begin{equation}
\begin{aligned}
c
%=&\frac{g_{pol-\gamma}^{2}g_{pol-ph}^{2}g_{pol-\gamma}^{2}}{[(\hbar\omega_{pol+}-\hbar\omega_{\gamma_{i}})^{2}+\Gamma_{pol+}^{2}][(\hbar\omega_{pol-}-\hbar\omega_{\gamma_{s}})^{2}+\Gamma_{pol-}^{2}]}
%\cdot\frac{[(E_{g}-\hbar\omega_{\gamma_{i}})^{2}+\Gamma_{e}^{2}][(E_{g}-\hbar\omega_{\gamma_{s}})^{2}+\Gamma_{e}^{2}]}{g_{e-\gamma}^{2}g_{e-ph}^{2}g_{e-\gamma}^{2}} \\
=\frac{g^{3}[(E_{g}-\hbar\omega_{\gamma_{i}})^{2}+\Gamma_{e}^{2}][(E_{g}-\hbar\omega_{\gamma_{s}})^{2}+\Gamma_{e}^{2}]}{[(\hbar\omega_{\mathrm{pol}+}-\hbar\omega_{\gamma_{i}})^{2}+\Gamma_{\mathrm{pol}+}^{2}][(\hbar\omega_{\mathrm{pol}-}-\hbar\omega_{\gamma_{s}})^{2}+\Gamma_{\mathrm{pol}-}^{2}]},
\end{aligned}
\end{equation}
where $g$ is defined as a normalized effective coupling term, which incorporates the coupling contributions from both the polariton scattering pathway and the ordinary Raman pathway, the final parameters requiring fitting are reduced to only three, $g,\ \Gamma_{\mathrm{pol}+},\ \Gamma_{\mathrm{pol}-}$.

\begin{figure}[htbp]
\centering
\includegraphics[width=0.45\textwidth]{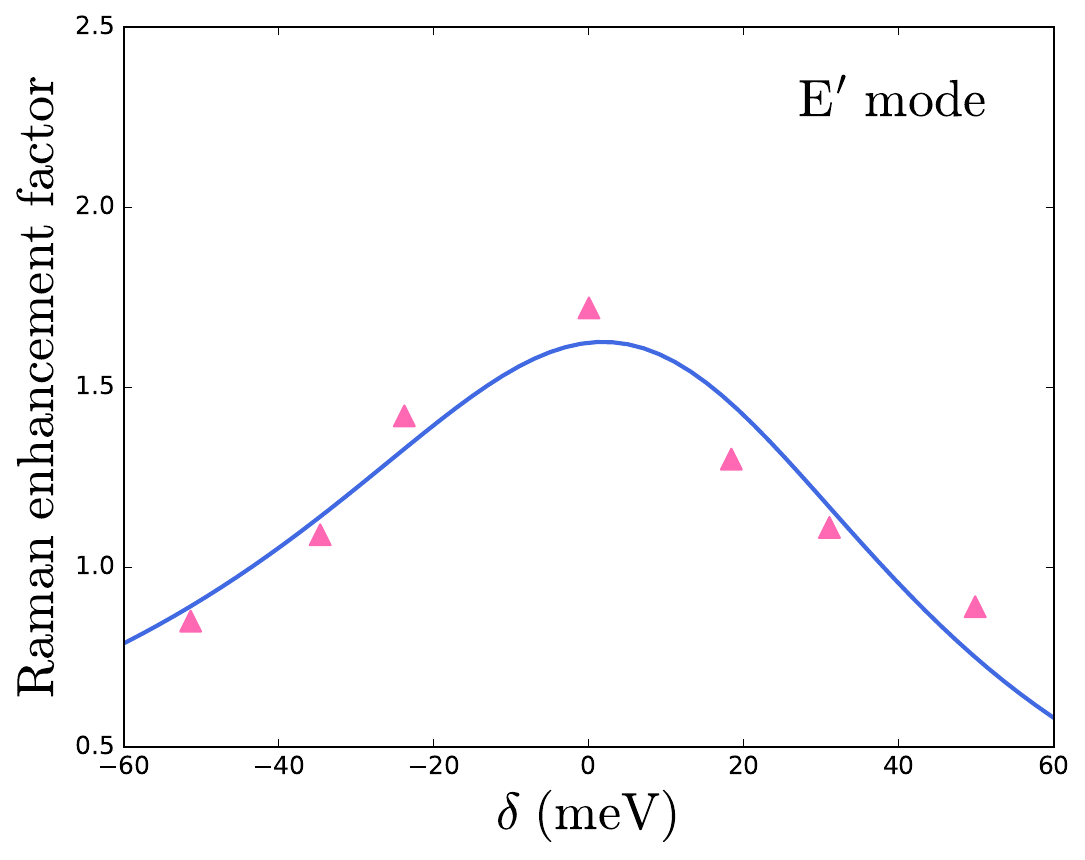}
\includegraphics[width=0.45\textwidth]{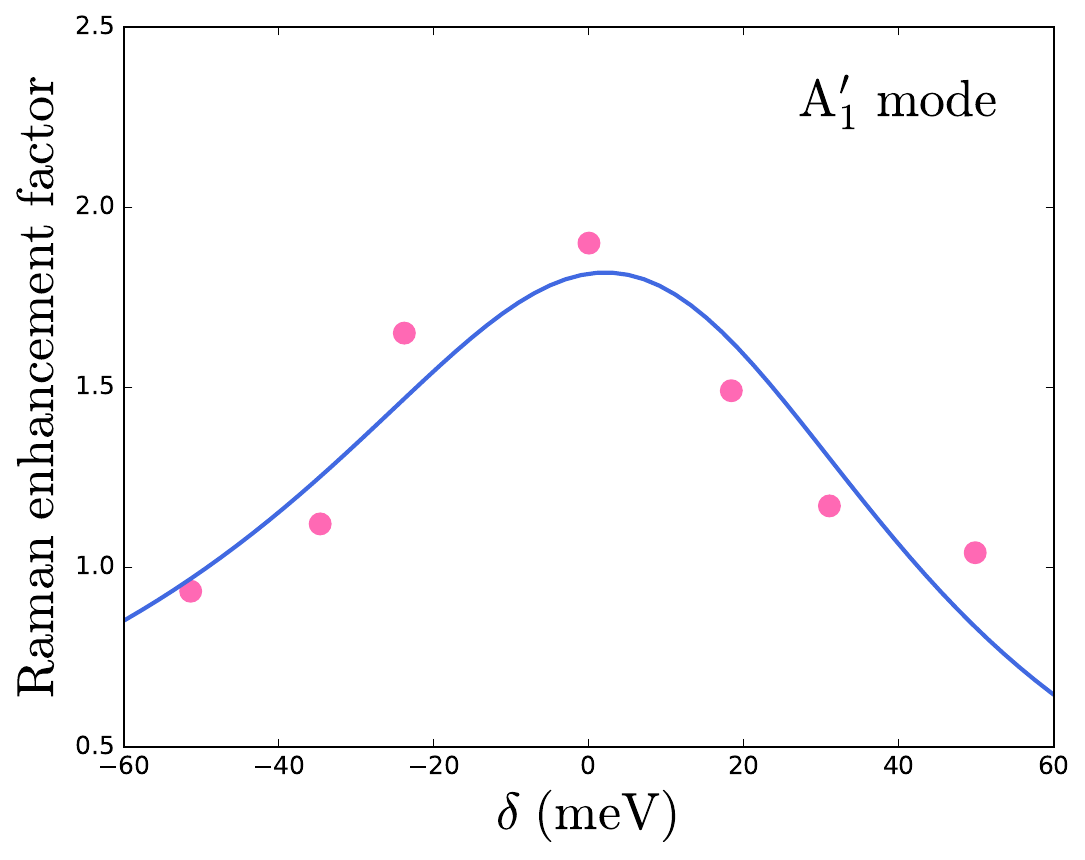}
\includegraphics[width=0.45\textwidth]{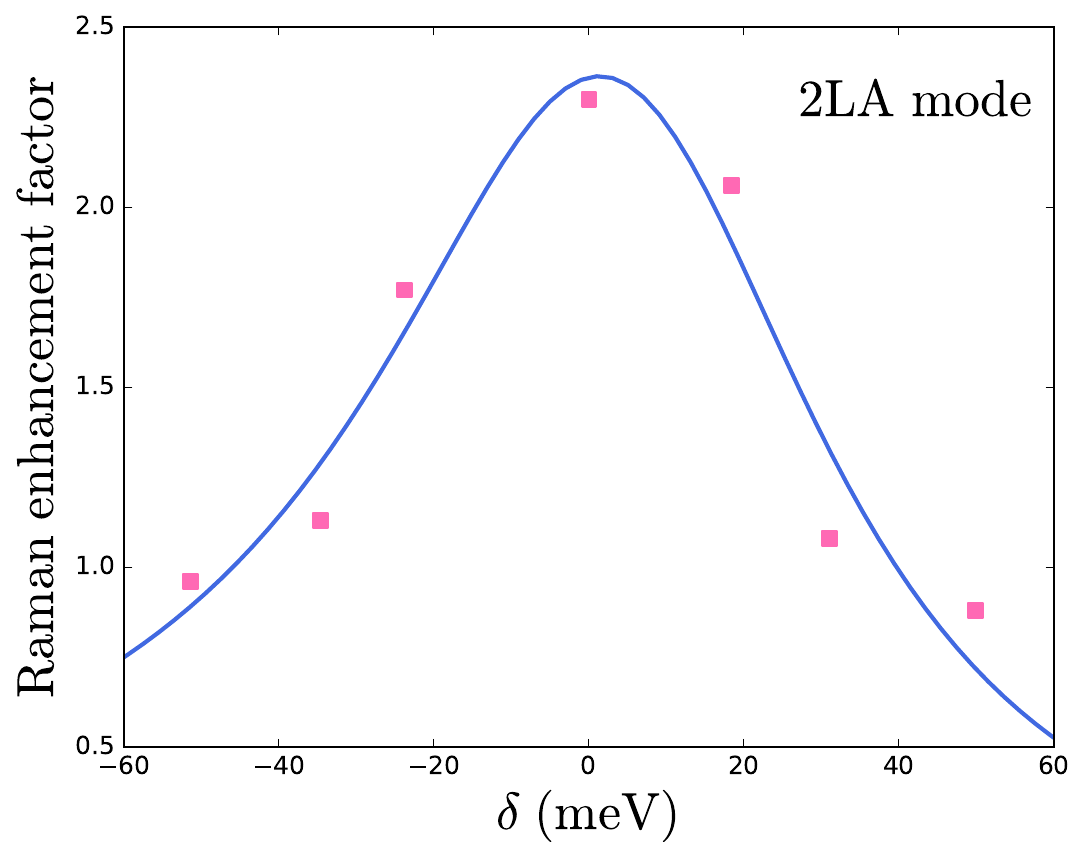}
\caption{\justifying Raman enhancement factors of phonon modes.
The phonon peak of the $\mathrm{E'}$ mode is at $388\ \mathrm{cm^{-1}}$, corresponding to an energy of $48.11\ \mathrm{meV}$. Fitting yields the widths of the upper and lower branch polaritons as $25.93\ \mathrm{meV}$ and $231.44\ \mathrm{meV}$, respectively, with $g=0.2$.
For the $\mathrm{A_{1}'}$ mode ($408\ \mathrm{cm^{-1}}$, $50.59\ \mathrm{meV}$), the widths of its upper and lower branch polaritons are $25.58\ \mathrm{meV}$ and $206.42\ \mathrm{meV}$, ($g = 0.18$). For the $\mathrm{2LA}$ mode ($460\ \mathrm{cm^{-1}}$, $57.04\ \mathrm{meV}$), the widths of its upper and lower branch polaritons are $18.41\ \mathrm{meV}$ and $175.85\ \mathrm{meV}$ ($g = 0.1$).}
\label{fig:raman_fit}
\end{figure}

We use the simulated Raman enhancement factors of three typical phonon modes in monolayer $\mathrm{MoS_{2}}$ to fit the effective coupling coefficient and the widths of the upper and lower polaritons in the expression.
By performing Lorentzian fitting on the Raman scattering spectra of monolayer $\mathrm{MoS_2}$, the center peak positions of various phonon modes were precisely extracted. We focus on three representative phonon modes: the $\mathrm{E'}$ mode (located at $388\ \mathrm{cm^{-1}}$, $48.11\ \mathrm{meV}$), the $\mathrm{A_{1}'}$ mode (located at $408\ \mathrm{cm^{-1}}$, $50.59\ \mathrm{meV}$), and the $\mathrm{2LA}$ mode (located at $460\ \mathrm{cm^{-1}}$, $57.04\ \mathrm{meV}$).
The incident photon wavelength is $633\ \mathrm{nm}$, corresponding to an energy of $1.96\ \mathrm{eV}$.
Figure~\ref{fig:raman_fit} shows the fitting results of the Raman scattering enhancement factors for these three phonon modes. %Therein, the pink markers represent the simulated Raman enhancement factors obtained from numerical calculations under different detunings, while the blue solid line is the fitting curve based on the analytical expression of the Raman enhancement factor. Through this fitting, we successfully extracted the polariton linewidths and the effective coupling strength coefficients for the different phonon modes.

%\clearpage

By analyzing the variation of the Raman enhancement factor with the detuning, we find that when the detuning is zero, the incident light energy approaches the upper-branch polariton energy, leading to resonance and a maximum Raman enhancement. The fitting results yield an effective coupling factor on the order of 0.1. Notably, the fitted width of the upper-branch polariton is significantly smaller than that of the lower-branch polariton, with a difference of nearly an order of magnitude. This phenomenon will be explained in the next chapter.

\section{plexciton width}
While formulating the total response function of the system, the effective dielectric function $\epsilon_{\mathrm{eff}}(q,\omega)$ was derived. This function can be expressd in terms of the Coulomb potential $v_{\mathrm{c}}$, the exciton potential $v_{\mathrm{ex}}$, and the density response function $\tilde{\chi}_{nn}(q,\omega)$. When dissipation is taken into account, the response function must be treated as a complex function by incorporating its imaginary part. Consequently, the finite width of the exciton must also be considered, with the exciton potential taking the form $v_{\mathrm{ex}}=g^{2}/(\omega-\omega_{\mathrm{ex}}+i\eta)$~\cite{antonius2022theory,berghauser2014analytical}, the effective dielectric function $\epsilon_{\mathrm{eff}}(q,\omega)$ consists of both real and imaginary parts,
%In the following section, we derive the explicit expressions for both the real and imaginary parts of the effective dielectric function,
\begin{equation}
\epsilon_{\mathrm{eff}}(q,\omega)=\epsilon_{\mathrm{eff}}^{r}(q,\omega)+i\epsilon_{\mathrm{eff}}^{i}(q,\omega).
\end{equation}
The real and imaginary parts of the effective dielectric function can be expressed in terms of the real and imaginary parts of the potentials and the response function
\begin{equation}
\begin{aligned}
&\epsilon_{\mathrm{eff}}^{r}(q,\omega)
=\dfrac{1}{2}-\dfrac{1}{2}(v^{r}\tilde{\chi}_{nn}^{r}-v^{i}\tilde{\chi}_{nn}^{i}),\\
&\epsilon_{\mathrm{eff}}^{i}(q,\omega)
=-\dfrac{1}{2}(v^{r}\tilde{\chi}_{nn}^{i}+v^{i}\tilde{\chi}_{nn}^{r}).
\end{aligned}
\end{equation}
Here, $v$ represents the sum of the Coulomb potential and the exciton potential, $v=v_{\mathrm{c}}+v_{\mathrm{ex}}$. It's real and imaginary parts are expressed as
\begin{equation}
v^{r}=\dfrac{N_{e}}{\tilde{q}}+\dfrac{N_{g''}\Delta}{\Delta^{2}+N_{\eta}^{2}},\
v^{i}=-\dfrac{N_{g'}}{\Delta^{2}+N_{\eta}^{2}},
\end{equation}
where we have defined the normalization constant,
\begin{equation}
\begin{aligned}
&N_{e}=\dfrac{2\pi e^{2}}{k_{F}},\
N_{g''}=\dfrac{\hbar g^{2}}{\epsilon_{F}},\
N_{\eta}=\dfrac{\hbar\eta}{\epsilon_{F}},\\
&N_{g'}=\dfrac{\hbar^{2}g^{2}\eta}{\epsilon_{F}^{2}}=N_{g''}N_{\eta},\
\Delta=\tilde{\omega}-\tilde{\omega}_{ex}.
\end{aligned}
\end{equation}

\begin{figure}[h]
\centering
\includegraphics[width=0.45\textwidth]{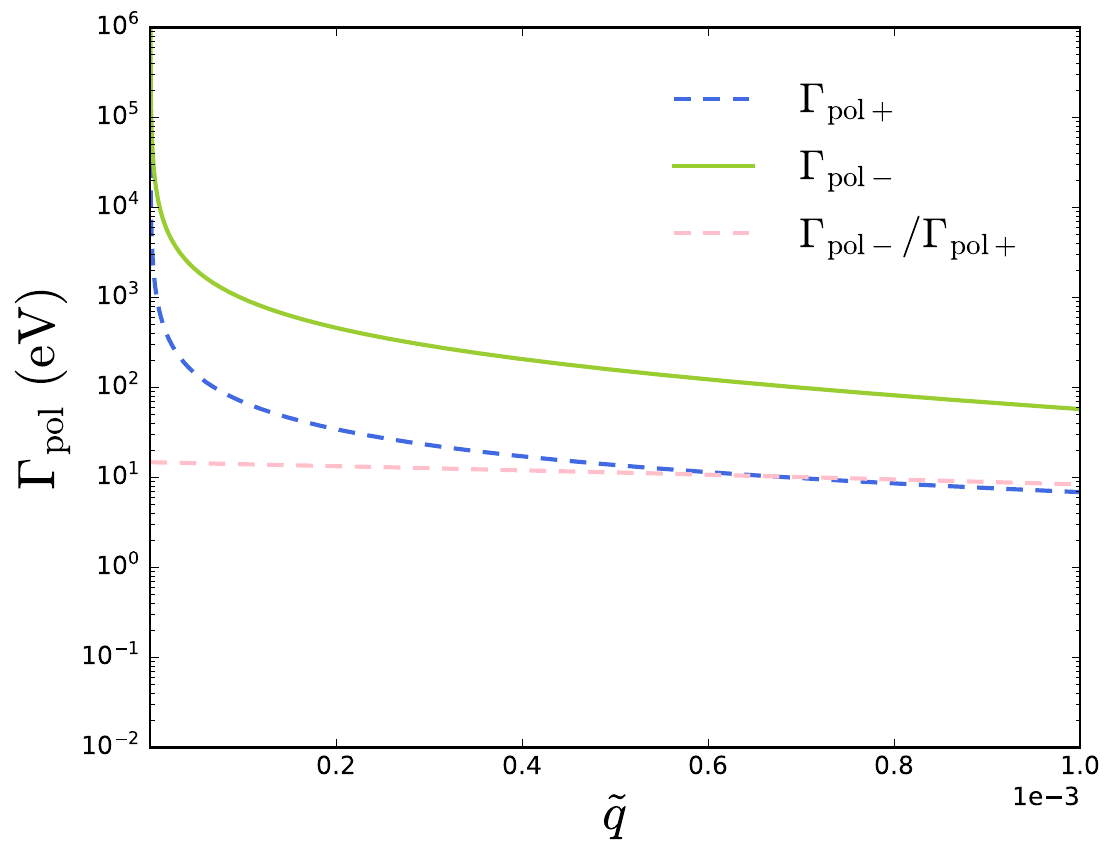}
\caption{\justifying Width of Upper and Lower Branch Polarons. When the imaginary part of the response function is taken into account, the widths of the polaritons are as shown in the figure. The widths of the upper and lower polaritons differ by a factor of ten. The width of the upper branch polariton is much smaller than that of the lower branch polariton.}
\label{fig:gamma_pol}
\end{figure}

The function $\chi_{0}(q,\omega)$ is a complex function consisting of both real and imaginary parts. In the long-wavelength limit $q\rightarrow 0$,
\begin{equation}
\begin{aligned}
&\tilde{\chi}_{nn}^{r}(q,\omega)
=\dfrac{2n}{\epsilon_{F}}\dfrac{\tilde{q}^{2}}{\tilde{\omega}^{2}}\left(1+a_{d}\dfrac{4\tilde{q}^{2}}{\tilde{\omega}^{2}}\right),\\
&\tilde{\chi}_{nn}^{i}(q,\omega)
=-\dfrac{N_{\sigma}}{\sqrt{4-\tilde{q}^{2}}}\dfrac{\tilde{\omega}}{\tilde{q}},
\end{aligned}
\end{equation}
where $N_{\sigma}$ is the density of states at the Fermi surface, expressed as $N_{\sigma}=m/(2\pi\hbar^{2})$~\cite{giuliani2008quantum}. The zeros of the effective dielectric function $\epsilon_{\text{eff}}(q,\omega)$ give the dispersion of the polariton. If dissipation is not considered, the zeros of the real part of the effective dielectric function $\epsilon_{\text{eff}}^{r}(q,\omega)$ yield the real polariton dispersion $\omega_{\text{pol}\pm}$ as discussed above.
When dissipation is taken into account, the polariton frequency becomes complex, $\omega_{\text{pol}}-i\Gamma$, with $\Gamma\neq 0$, satisfying $\epsilon_{\text{eff}}(q,\omega)=0$.
Expanding the effective dielectric function around the real frequency $\omega_{\text{pol}}$,
\begin{equation}
\begin{aligned}
\epsilon_{\mathrm{eff}}(q,\omega_{\text{pol}}-i\Gamma)
=\epsilon_{\mathrm{eff}}(q,\omega_{\text{pol}})+\left.\dfrac{\partial\epsilon_{\mathrm{eff}}}{\partial\omega}\right|_{\omega_{\text{pol}}}(-i\Gamma)=0,
%=\epsilon_{eff}^{r}(q,\omega_{pol})+i\epsilon_{eff}^{i}(q,\omega_{pol})+\left.\left(\dfrac{\partial\epsilon_{eff}^{r}}{\partial\omega}+i\dfrac{\partial\epsilon_{eff}^{i}}{\partial\omega}\right)\right|_{\omega_{pol}}(-i\Gamma)=0,
\end{aligned}
\end{equation}

we express the real and imaginary parts of the equation,
\begin{equation}
\begin{aligned}
&\left[\epsilon_{\mathrm{eff}}^{r}(q,\omega_{\text{pol}})+\left.\left(\dfrac{\partial\epsilon_{\mathrm{eff}}^{i}}{\partial\omega}\right)\right|_{\omega_{\text{pol}}}\Gamma\right] \\
&+i\left[\epsilon_{\mathrm{eff}}^{i}(q,\omega_{\text{pol}})-\left.\left(\dfrac{\partial\epsilon_{\mathrm{eff}}^{r}}{\partial\omega}\right)\right|_{\omega_{\text{pol}}}\Gamma\right]=0.
\end{aligned}
\end{equation}
When only the real part of the dielectric function is considered, the polariton angular frequency satisfies $\epsilon_{\mathrm{eff}}^{r}(q, \omega_{\mathrm{pol}}) = 0$.
Assuming the rate of change of the imaginary part of the dielectric function is small, the terms within the first curly bracket in the above expression vanish. We thus obtain the expression for the polariton width:
\begin{equation}
\begin{aligned}
\Gamma
=&\dfrac{\epsilon_{\mathrm{eff}}^{i}(q,\omega_{\text{pol}})}{\left.\left(\dfrac{\partial\epsilon_{\mathrm{eff}}^{r}}{\partial\omega}\right)\right|_{\omega_{\text{pol}}}} \\
=&\dfrac{v^{r}\tilde{\chi}_{nn}^{i}+v^{i}\tilde{\chi}_{nn}^{r}}{\dfrac{\partial v^{r}}{\partial\omega}\tilde{\chi}_{nn}^{r}+v^{r}\dfrac{\partial\tilde{\chi}_{nn}^{r}}{\partial\omega}-\dfrac{\partial v^{i}}{\partial\omega}\tilde{\chi}_{nn}^{i}-v^{i}\dfrac{\partial\tilde{\chi}_{nn}^{i}}{\partial\omega}}.
\end{aligned}
\end{equation}
Based on this expression, we substitute the dispersion relations for the upper and lower polariton branches, $\omega_{\mathrm{pol\pm}}$, into the equation to obtain the respective polariton widths, as shown in Figure~\ref{fig:gamma_pol}.

It can be clearly seen from the figure that the width of the lower branch polariton is much larger than that of the upper branch polariton, with the difference between the two widths being nearly tenfold. In fact, in the calculations of this chapter, we consider the plasmon width within the electron-hole continuum to compute the polariton width. Usually, the resonant wavevector of the polariton does not lie within the electron-hole continuum. Although, from the perspective of the dispersion relation, the resonant wavevector of the polariton is located outside the theoretical boundary of the electron-hole continuum, this does not mean that the polariton is unaffected by damping mechanisms. The intrinsic dissipation of the electron-hole continuum broadens the boundary of the continuum. The lower polariton, having lower energy and lying in the high-loss region near the continuum edge, experiences residual damping much stronger than that of the higher-energy upper branch polariton. This is one possible reason for the highly asymmetric linewidths of the upper and lower polariton branches. Furthermore, the lower polariton branch resides in a lower-energy region where the density of single-particle excited states is much higher than in the higher-energy region. This extremely high density of states creates an environment in which the lower polariton readily undergoes scattering and energy exchange with other excitation modes in the background through strong coupling effects, leading to a dramatic increase in its dissipation rate and thus a much larger linewidth compared to the upper branch.

\section{conclusion}
In our work, we derive the dispersion relation of plasmons in metals based on linear response theory. For a simple experimental system---such as a plasmonic cavity composed of gold nanoparticles and semiconductor materials---we clarify that varying the size of the metal particles effectively shifts the resonance wave vector between the plasmons and the incident light, thereby determining a specific plasmonic mode involved in the coupling.
By calculating the effective dielectric function of the entire coupled system, we derive the dispersion relation for plasmon-exciton polaritons, yielding both the upper and lower polariton branches. We describe the scattering amplitude and intensity of polariton resonant Raman scattering at the quantum level. In comparison with conventional Raman scattering, we provide an expression for the polariton Raman enhancement factor, which aligns well with experimental results.
The fitting results for the Raman enhancement factor indicate that the upper and lower polariton branches possess different linewidths.
By incorporating the imaginary part of the dielectric function into our expressions to account for the system's dissipation, we offer an insightful perspective for understanding the significant asymmetry in the linewidths of the upper and lower branch polaritons.

\begin{acknowledgments}
This work is supported by National Natural Science Foundation of China, General Program 122741381004424, Henan Academy of Sciences, Zhongyuan Academician Research Fund 20261724001, National Natural Science Foundation of China, Science Center Program 52488301, and Academic Divisions of Chinese Academy of Sciences, Grant 2021-SL01-A-028.
\end{acknowledgments}

\appendix
\section{linear response theory and dielectric function}
According to linear response theory, the charge density response and the external perturbation potential satisfy a linear relation:
\begin{equation}
n_{\mathrm{ind}}(q,\omega)=\chi_{nn}(q,\omega)V_{\mathrm{ext}}(q,\omega).
\label{eq_ind1}
\end{equation}
where $\chi_{nn}$ is the density-density response function. The Coulomb potential generated by the induced charge is
\begin{equation}
V_{\mathrm{ind}}(q,\omega)=v_{\mathrm{c}}\chi_{nn}(q,\omega)V_{\mathrm{ext}}(q,\omega).
\end{equation}
The sum of the external potential and the induced Coulomb potential constitutes the screened scalar potential, which can be measured by a test charge:
\begin{equation}
\begin{aligned}
V_{\mathrm{sc}}(q,\omega)=&V_{\mathrm{ext}}(q,\omega)+V_{\mathrm{ind}}(q,\omega) \\
=&[1+v_{\mathrm{c}}\chi_{nn}(q,\omega)]V_{\mathrm{ext}}(q,\omega).
\label{eq_scp}
\end{aligned}
\end{equation}
We further define the proper density-density response function $\tilde{\chi}_{nn}(q,\omega)$, which describes the response of the charge density to the screened potential:
\begin{equation}
\begin{aligned}
n_{\mathrm{ind}}(q,\omega)
=&\tilde{\chi}_{nn}(q,\omega)V_{\mathrm{sc}}(q,\omega) \\
=&\tilde{\chi}_{nn}(q,\omega)[1+v_{\mathrm{c}}\chi_{nn}(q,\omega)]V_{\mathrm{ext}}(q,\omega),
\label{eq_ind2}
\end{aligned}
\end{equation}
Comparing Eq. \ref{eq_ind1} with Eq. \ref{eq_ind2}, we obtain the relationship between the two response functions:
\begin{equation}
\chi_{nn}(q,\omega)
=\dfrac{\tilde{\chi}_{nn}(q,\omega)}{1-v_{\mathrm{c}}\tilde{\chi}_{nn}(q,\omega)}.
\label{eq_chinn}
\end{equation}
Substituting into \ref{eq_scp} yields the relation between the screened potential and the external potential:
\begin{equation}
\begin{aligned}
V_{\mathrm{sc}}(q,\omega)
=&\left[ 1+\dfrac{v_{\mathrm{c}}\tilde{\chi}_{nn}(q,\omega)}{1-v_{\mathrm{c}}\tilde{\chi}_{nn}(q,\omega)}\right] V_{\mathrm{ext}}(q,\omega) \\
=&\dfrac{V_{\mathrm{ext}}(q,\omega)}{1-v_{\mathrm{c}}\tilde{\chi}_{nn}(q,\omega)}
=\dfrac{V_{\mathrm{ext}}(q,\omega)}{\epsilon(q,\omega)},
\end{aligned}
\end{equation}
from which the dielectric function is defined as:
\begin{equation}
\epsilon(q,\omega)=1-v_{\mathrm{c}}\tilde{\chi}_{nn}(q,\omega).
\end{equation}
Thus, the screened potential differs from the external potential by a dielectric factor. Meanwhile, from \ref{eq_chinn} we also obtain:
\begin{equation}
\chi_{nn}(q,\omega)
=\dfrac{\tilde{\chi}_{nn}(q,\omega)}{\epsilon(q,\omega)}.
\end{equation}
the density-density response function and the proper density-density response function also differ by the same dielectric function.

\section{plasmon dispersion}
The zeros of the dielectric function $\epsilon(q,\omega)$ yield the dispersion relation of plasmon. Solving the equation $\epsilon(q,\omega)=0$ is equivalent to solving $v_{\mathrm{c}}(q)\tilde{\chi}_{nn}(q,\omega)=1$. By substituting the Coulomb potential and the proper density-density response function under the RPA into the equation. Calculate
\begin{equation}
\dfrac{2\pi e^{2}}{q}\dfrac{nq^{2}}{m\omega^{2}}\left(1+a_{d}\dfrac{q^{2}v_{F}^{2}}{\omega^{2}}\right)
%=\dfrac{2\pi ne^{2}q}{m\omega^{2}}\left(1+a_{d}\dfrac{q^{2}v_{F}^{2}}{\omega^{2}}\right)
=\dfrac{\omega_{p}^{2}}{\omega^{2}}\left(1+a_{d}\dfrac{q^{2}v_{F}^{2}}{\omega^{2}}\right)=1,
\label{eq:plasmon_eq}
\end{equation}
where, $\omega_{p}=\sqrt{\dfrac{2\pi ne^{2}q}{m}}$. This is a quartic equation with respect to $\omega$:
\begin{equation}
\omega^{4}-\omega_{p}^{2}\omega^{2}-a_{d}q^{2}v_{F}^{2}\omega_{p}^{2}=0.
\end{equation}
Let $\Omega=\omega^{2}$, the equation becomes
\begin{equation}
\Omega^{2}-\omega_{p}^{2}\Omega-a_{d}q^{2}v_{F}^{2}\omega_{p}^{2}=0,
\end{equation}
the solution to the equation is given by
\begin{equation}
\Omega
%=\dfrac{1}{2}\left(\omega_{p}^{2}\pm\sqrt{\omega_{p}^{4}+4a_{d}q^{2}v_{F}^{2}\omega_{p}^{2}}\right)
=\dfrac{\omega_{p}^{2}}{2}\left(1\pm\sqrt{1+4a_{d}\dfrac{q^{2}v_{F}^{2}}{\omega_{p}^{2}}}\right),
\end{equation}
since $\Omega=\omega^{2}>0$, we take the positive root of the equation. In the long-wave length limit $q\rightarrow 0$,
\begin{equation}
\Omega
\simeq\dfrac{\omega_{p}^{2}}{2}\left(1+1+\dfrac{4a_{d}}{2}\dfrac{q^{2}v_{F}^{2}}{\omega_{p}^{2}}\right)
%=\omega_{p}^{2}\left(1+a_{d}\dfrac{q^{2}v_{F}^{2}}{\omega_{p}^{2}}\right)
=\omega_{p}^{2}+a_{d}q^{2}v_{F}^{2}.
\end{equation}
The plasmon dispersion relation reads
\begin{equation}
\omega_{\mathrm{pl}}=\sqrt{\Omega}
%=\sqrt{\omega_{p}^{2}+a_{d}q^{2}v_{F}^{2}}
%=\omega_{p}\sqrt{1+\dfrac{a_{d}q^{2}v_{F}^{2}}{\omega_{p}^{2}}}
\simeq\omega_{p}\left(1+\dfrac{1}{2}\dfrac{a_{d}q^{2}v_{F}^{2}}{\omega_{p}^{2}}\right)
=\omega_{p}\left(1+a_{d}\dfrac{q}{\kappa_{2}}\right),
\label{eq:w_pl_app}
\end{equation}
where, $\kappa_{2}=\dfrac{2\omega_{p}^{2}}{qv_{F}^{2}}=\dfrac{4\pi ne^{2}}{mv_{F}^{2}}$.

Furthermore, the plasmon dispersion relation can be normalized by the Fermi wavefactor $k_{F}$ and Fermin energy $\epsilon_{F}$. We define the normalized variables as
\begin{equation}
\tilde{\omega}_{\mathrm{pl}}=\dfrac{\hbar\omega_{\mathrm{pl}}}{\epsilon_{F}},\
\tilde{q}=\dfrac{q}{k_{F}}.
\end{equation}
On a clean noble metal surface, the surface-state electrons can be regarded as a nearly ideal two-dimensional free electron gas. Consequently, the relationship between the Fermi wave vector $k_{F}$ and the electron areal density $n$ follows $k_{F}=\sqrt{2\pi n}$~\cite{kevan1987high,petersen1998direct,reinert2001direct,berland2012response}.
Conversely, the charge density can be expressed in terms of the Fermi wavevector as $n=\dfrac{k_{F}^{2}}{2\pi}$. We then define the characterisitic frequency as
\begin{equation}
\omega_{F}
=\omega_{p}(k_{F})
=\sqrt{\dfrac{2\pi ne^{2}k_{F}}{m}}
=\sqrt{\dfrac{k_{F}^{3}e^{2}}{m}},
\end{equation}
then,
\begin{equation}
\omega_{p}(q)
=\sqrt{\dfrac{2\pi ne^{2}q}{m}}
=\sqrt{\dfrac{k_{F}^{3}e^{2}\tilde{q}}{m}}
=\omega_{F}\sqrt{\tilde{q}}.
\end{equation}
Multiplying both sides of Eq.~(\ref{eq:w_pl_app}) by $\hbar/\epsilon_{F}$, then
\begin{equation}
\dfrac{\hbar\omega_{\mathrm{pl}}}{\epsilon_{F}}
=\dfrac{\hbar\omega_{p}}{\epsilon_{F}}\left(1+a_{d}\dfrac{q}{\kappa_{2}}\right)
=\dfrac{\hbar\omega_{F}}{\epsilon_{F}}\sqrt{\tilde{q}}\left(1+a_{d}\dfrac{\tilde{q}k_{F}}{\kappa_{2}}\right),
\end{equation}
therefore, the normalized dispersion relation is
\begin{equation}
\tilde{\omega}_{\mathrm{pl}}
=\tilde{\omega}_{F}\sqrt{\tilde{q}}\left(1+a_{d}\dfrac{\tilde{q}}{\tilde{\kappa}_{2}}\right),
\end{equation}
here, we define two normalization constants:
\begin{equation}
\tilde{\omega}_{F}=\dfrac{\hbar\omega_{F}}{\epsilon_{F}},\
\tilde{\kappa}_{2}=\dfrac{\kappa_{2}}{k_{F}}.
\end{equation}
Note that there exists the relationship between the Fermi wavevector $k_{F}$, Fermi velocity $v_{F}$, and the Fermi energy $\epsilon_{F}$,
\begin{equation}
\epsilon_{F}=\dfrac{\hbar^{2}k_{F}^{2}}{2m},\
v_{F}=\dfrac{\hbar k_{F}}{m},\
v_{F}^{2}%=\dfrac{\hbar^{2}k_{F}^{2}}{m^{2}}=\dfrac{2m\epsilon_{F}}{m^{2}}
=\dfrac{2\epsilon_{F}}{m}.
\end{equation}
The two coefficients can be further simplified,
\begin{equation}
\begin{aligned}
&\tilde{\omega}_{F}
=\dfrac{\hbar\omega_{F}}{\epsilon_{F}}
=\hbar\sqrt{\dfrac{k_{F}^{3}e^{2}}{m}}\dfrac{2m}{\hbar^{2}k_{F}^{2}}
=\dfrac{2e}{\hbar}\sqrt{\dfrac{m}{k_{F}}}
,\\
&\tilde{\kappa}_{2}=\dfrac{\kappa_{2}}{k_{F}}
=\dfrac{4\pi ne^{2}}{mv_{F}^{2}}\dfrac{1}{k_{F}}
=\dfrac{4\pi e^{2}}{mk_{F}}\dfrac{k_{F}^{2}}{2\pi}\dfrac{m^{2}}{\hbar^{2}k_{F}^{2}}
=\dfrac{2me^{2}}{\hbar^{2}k_{F}}.
\end{aligned}
\end{equation}

\section{plexciton dispersion}
The zeros of the effective dielectric function $\epsilon_{\mathrm{eff}}(q,\omega)$ yield the dispersion relation of the polariton. Solving the equation $\epsilon_{\mathrm{eff}}(q,\omega)=0$ is equivalent to solving:
\begin{equation}
(v_{\mathrm{c}}+v_{\mathrm{ex}})\tilde{\chi}_{nn}(q,\omega)=1.
\end{equation}
calculate:
\begin{equation}
\begin{aligned}
&\left(\dfrac{2\pi e^{2}}{q}+\dfrac{g^{2}}{\omega-\omega_{\mathrm{ex}}}\right)\dfrac{nq^{2}}{m\omega^{2}}\left(1+a_{d}\dfrac{q^{2}v_{F}^{2}}{\omega^{2}}\right) \\
&=\left[\dfrac{2\pi ne^{2}q}{m\omega^{2}}+\dfrac{ng^{2}q^{2}}{m\omega^{2}(\omega-\omega_{\mathrm{ex}})}\right]\left(1+a_{d}\dfrac{q^{2}v_{F}^{2}}{\omega^{2}}\right) \\
&=\left[\dfrac{\omega_{p}^{2}}{\omega^{2}}+\dfrac{ng^{2}q^{2}}{m\omega^{2}(\omega-\omega_{\mathrm{ex}})}\right]\left(1+a_{d}\dfrac{q^{2}v_{F}^{2}}{\omega^{2}}\right)=1.
\end{aligned}
\end{equation}

The plasmon angular frequency $\omega_{\mathrm{pl}}$ satisfies Eq.~(\ref{eq:plasmon_eq}),
\begin{equation}
\dfrac{\omega_{p}^{2}}{\omega_{\mathrm{pl}}^{2}}\left(1+a_{d}\dfrac{q^{2}v_{F}^{2}}{\omega_{\mathrm{pl}}^{2}}\right)=1,
\end{equation}
Using this relationship, $\omega_{p}$ can be expressed in terms of $\omega_{\mathrm{pl}}$,
\begin{equation}
\omega_{p}^{2}=\omega_{\mathrm{pl}}^{2}\left(1+a_{d}\dfrac{q^{2}v_{F}^{2}}{\omega_{\mathrm{pl}}^{2}}\right)^{-1}.
\end{equation}

Substituting the above expression into the equation and neglecting higher-order terms $O(q^{2})$, we obtain
\begin{equation}
\begin{aligned}
&\left[\dfrac{\omega_{\mathrm{pl}}^{2}}{\omega^{2}}\left(1+a_{d}\dfrac{q^{2}v_{F}^{2}}{\omega_{\mathrm{\mathrm{pl}}}^{2}}\right)^{-1}+\dfrac{ng^{2}q^{2}}{m\omega^{2}(\omega-\omega_{\mathrm{ex}})}\right]\left(1+a_{d}\dfrac{q^{2}v_{F}^{2}}{\omega^{2}}\right) \\
=&\dfrac{\omega_{\mathrm{\mathrm{pl}}}^{2}}{\omega^{2}}+\dfrac{ng^{2}q^{2}}{m\omega^{2}(\omega-\omega_{\mathrm{ex}})}=1,
\end{aligned}
\end{equation}
this equation is a cubic equation in terms of $\omega$:
\begin{equation}
\omega^{3}-\omega_{\mathrm{ex}}\omega^{2}-\omega_{\mathrm{pl}}^{2}\omega+\omega_{\mathrm{pl}}^{2}\omega_{\mathrm{ex}}-\dfrac{ng^{2}q^{2}}{m}=0.
\end{equation}
Considering the polariton frequency $\omega$ near the exciton frequency $\omega_{\mathrm{ex}}$. Let $\omega=\omega_{\mathrm{ex}}+\delta$, substituting this into above equation and neglecting high-order terms of $\delta^{3}$, we obtain
\begin{equation}
\begin{aligned}
&(\omega_{\mathrm{ex}}+\delta)^{3}-\omega_{\mathrm{ex}}(\omega_{\mathrm{ex}}+\delta)^{2}-\omega_{\mathrm{pl}}^{2}(\omega_{\mathrm{ex}}+\delta)+\omega_{\mathrm{pl}}^{2}\omega_{\mathrm{ex}} \\
&-\dfrac{ng^{2}q^{2}}{m}\\
=&\omega_{\mathrm{ex}}^{3}+3\omega_{\mathrm{ex}}^{2}\delta+3\omega_{\mathrm{ex}}\delta^{2}+\delta^{3}-\omega_{\mathrm{ex}}^{3}-2\omega_{\mathrm{ex}}^{2}\delta-\omega_{\mathrm{ex}}\delta^{2} \\
&-\omega_{\mathrm{pl}}^{2}\omega_{\mathrm{ex}}-\omega_{\mathrm{pl}}^{2}\delta+\omega_{\mathrm{pl}}^{2}\omega_{\mathrm{ex}}-\dfrac{ng^{2}q^{2}}{m}\\
=&\delta^{3}+2\omega_{\mathrm{ex}}\delta^{2}+(\omega_{\mathrm{ex}}^{2}-\omega_{\mathrm{pl}}^{2})\delta-\dfrac{ng^{2}q^{2}}{m}\\
\approx &2\omega_{\mathrm{ex}}\delta^{2}+(\omega_{\mathrm{ex}}^{2}-\omega_{\mathrm{pl}}^{2})\delta-\dfrac{ng^{2}q^{2}}{m},
\end{aligned}
\end{equation}
the solution of the equation is
\begin{equation}
\begin{aligned}
\delta
=&\dfrac{1}{4\omega_{\mathrm{ex}}}\left(-(\omega_{\mathrm{ex}}^{2}-\omega_{\mathrm{pl}}^{2})\pm\sqrt{(\omega_{\mathrm{ex}}^{2}-\omega_{\mathrm{pl}}^{2})^{2}+8\omega_{\mathrm{ex}}\dfrac{ng^{2}q^{2}}{m}}\right) \\
=&-\dfrac{\omega_{\mathrm{ex}}}{4}+\dfrac{\omega_{\mathrm{pl}}^{2}}{4\omega_{\mathrm{ex}}}\pm\sqrt{\dfrac{(\omega_{\mathrm{ex}}^{2}-\omega_{\mathrm{pl}}^{2})^{2}}{16\omega_{\mathrm{ex}}^{2}}+\dfrac{ng^{2}q^{2}}{2m\omega_{\mathrm{ex}}}}.
\end{aligned}
\end{equation}

Thus, the dispersion relation of the plasmon-exciton polariton becomes
\begin{equation}
\begin{aligned}
\omega_{\mathrm{pol}\pm}
=&\omega_{\mathrm{ex}}+\delta \\
=&\dfrac{3\omega_{\mathrm{ex}}}{4}+\dfrac{\omega_{\mathrm{pl}}^{2}}{4\omega_{\mathrm{ex}}}\pm\sqrt{\dfrac{(\omega_{\mathrm{ex}}^{2}-\omega_{\mathrm{pl}}^{2})^{2}}{16\omega_{\mathrm{ex}}^{2}}+\dfrac{ng^{2}q^{2}}{2m\omega_{\mathrm{ex}}}}.
\end{aligned}
\end{equation}

\section{plexciton width}
The effective dielectric function can be expressed by Coulomb and exciton potential and response function as
\begin{equation}
\begin{aligned}
\epsilon_{\mathrm{eff}}(q,\omega)
&=\dfrac{1}{2}-\dfrac{1}{2}(v_{\mathrm{c}}+v_{\mathrm{ex}})\tilde{\chi}_{nn}(q,\omega)
=\dfrac{1}{2}-\dfrac{1}{2}v\tilde{\chi}_{nn}(q,\omega) \\
&=\dfrac{1}{2}-\dfrac{1}{2}(v^{r}+iv^{i})(\tilde{\chi}_{nn}^{r}+i\tilde{\chi}_{nn}^{i})\\
&=\dfrac{1}{2}[1-(v^{r}\tilde{\chi}_{nn}^{r}-v^{i}\tilde{\chi}_{nn}^{i})]-\dfrac{i}{2}(v^{r}\tilde{\chi}_{nn}^{i}+v^{i}\tilde{\chi}_{nn}^{r}),\\
\epsilon_{\mathrm{eff}}^{r}(q,\omega)
&=\dfrac{1}{2}-\dfrac{1}{2}(v^{r}\tilde{\chi}_{nn}^{r}-v^{i}\tilde{\chi}_{nn}^{i}),\\
\epsilon_{\mathrm{eff}}^{i}(q,\omega)
&=-\dfrac{1}{2}(v^{r}\tilde{\chi}_{nn}^{i}+v^{i}\tilde{\chi}_{nn}^{r}).
\end{aligned}
\end{equation}
The partial derivative of the dielectric function with respect to angular frequency is denoted as:
\begin{equation}
\begin{aligned}
&\dfrac{\partial\epsilon_{\mathrm{eff}}^{r}(q,\omega)}{\partial\omega}
=-\dfrac{1}{2}\left(\dfrac{\partial v^{r}}{\partial\omega}\tilde{\chi}_{nn}^{r}+v^{r}\dfrac{\partial\tilde{\chi}_{nn}^{r}}{\partial\omega}-\dfrac{\partial v^{i}}{\partial\omega}\tilde{\chi}_{nn}^{i}-v^{i}\dfrac{\tilde{\chi}_{nn}^{i}}{\partial\omega}\right),\\
&\dfrac{\partial\epsilon_{\mathrm{eff}}^{i}(q,\omega)}{\partial\omega}
=-\dfrac{1}{2}\left(\dfrac{\partial v^{r}}{\partial\omega}\tilde{\chi}_{nn}^{i}+v^{r}\dfrac{\partial\tilde{\chi}_{nn}^{i}}{\partial\omega}+\dfrac{\partial v^{i}}{\partial\omega}\tilde{\chi}_{nn}^{r}+v^{i}\dfrac{\tilde{\chi}_{nn}^{r}}{\partial\omega}\right).
\end{aligned}
\end{equation}
The normalized Coulomb and exciton potentials are given by
\begin{equation}
\begin{aligned}
v_{\mathrm{c}}&=\dfrac{2\pi e^{2}}{q},\\
v_{\mathrm{ex}}&
=\dfrac{g^{2}}{\omega-\omega_{\mathrm{ex}}+i\eta}
=\dfrac{g^{2}(\omega-\omega_{\mathrm{ex}})}{(\omega-\omega_{\mathrm{ex}})^{2}+\eta^{2}}-i\dfrac{g^{2}\eta}{(\omega-\omega_{\mathrm{ex}})^{2}+\eta^{2}},\\
v&=\left[\dfrac{2\pi e^{2}}{q}+\dfrac{g^{2}(\omega-\omega_{\mathrm{ex}})}{(\omega-\omega_{\mathrm{ex}})^{2}+\eta^{2}}\right]
-i\dfrac{g^{2}\eta}{(\omega-\omega_{\mathrm{ex}})^{2}+\eta^{2}},\\
v^{r}
&=\dfrac{2\pi e^{2}}{q}+\dfrac{g^{2}(\omega-\omega_{\mathrm{ex}})}{(\omega-\omega_{\mathrm{ex}})^{2}+\eta^{2}}
=\dfrac{2\pi e^{2}}{\tilde{q}k_{F}}+\dfrac{g^{2}(\dfrac{\epsilon_{F}}{\hbar}\tilde{\omega}-\omega_{\mathrm{ex}})}{(\dfrac{\epsilon_{F}}{\hbar}\tilde{\omega}-\omega_{\mathrm{ex}})^{2}+\eta^{2}}\\
&=\dfrac{2\pi e^{2}}{\tilde{q}k_{F}}+\dfrac{\dfrac{\hbar g^{2}}{\epsilon_{F}}(\tilde{\omega}-\tilde{\omega}_{\mathrm{ex}})}{(\tilde{\omega}-\tilde{\omega}_{\mathrm{ex}})^{2}+\dfrac{\hbar^{2}\eta^{2}}{\epsilon_{F}^{2}}}
=\dfrac{N_{e}}{\tilde{q}}+\dfrac{N_{g''}\Delta}{\Delta^{2}+N_{\eta}^{2}}, \\
v^{i}
&=-\dfrac{g^{2}\eta}{(\omega-\omega_{\mathrm{ex}})^{2}+\eta^{2}}
=-\dfrac{g^{2}\eta}{(\dfrac{\epsilon_{F}}{\hbar}\tilde{\omega}-\omega_{\mathrm{ex}})^{2}+\eta^{2}} \\
&=-\dfrac{\dfrac{\hbar^{2}g^{2}\eta}{\epsilon_{F}^{2}}}{(\tilde{\omega}-\tilde{\omega}_{\mathrm{ex}})^{2}+\dfrac{\hbar^{2}\eta^{2}}{\epsilon_{F}^{2}}}
=-\dfrac{N_{g'}}{\Delta^{2}+N_{\eta}^{2}},
\end{aligned}
\end{equation}
where we define:
\begin{equation}
\begin{aligned}
&N_{e}=\dfrac{2\pi e^{2}}{k_{F}},\
N_{g''}=\dfrac{\hbar g^{2}}{\epsilon_{F}},\
N_{\eta}=\dfrac{\hbar\eta}{\epsilon_{F}},\\
&N_{g'}=\dfrac{\hbar^{2}g^{2}\eta}{\epsilon_{F}^{2}}=N_{g''}N_{\eta},\
\Delta=\tilde{\omega}-\tilde{\omega}_{\mathrm{ex}}.
\end{aligned}
\end{equation}

Calculate the partial derivatives of the real and imaginary parts of the potential energy with respect to the angular frequency,
\begin{equation}
\begin{aligned}
\dfrac{\partial v^{r}}{\partial\omega}
&=\dfrac{\partial v^{r}}{\partial\tilde{\omega}}\dfrac{\partial\tilde{\omega}}{\partial\omega}
=\dfrac{\hbar}{\epsilon_{F}}\dfrac{\partial v^{r}}{\partial\tilde{\omega}}
=N_{\hbar}\dfrac{N_{g''}(\Delta^{2}+N_{\eta}^{2})-2N_{g''}\Delta^{2}}{(\Delta^{2}+N_{\eta}^{2})^{2}}\\
&=\dfrac{N_{\hbar}N_{g''}(N_{\eta}^{2}-\Delta^{2})}{(\Delta^{2}+N_{\eta}^{2})^{2}}
=\dfrac{N_{g}^{2}(N_{\eta}^{2}-\Delta^{2})}{(\Delta^{2}+N_{\eta}^{2})^{2}}, \\
\dfrac{\partial v^{i}}{\partial\omega}
&=\dfrac{\partial v^{i}}{\partial\tilde{\omega}}\dfrac{\partial\tilde{\omega}}{\partial\omega}
=\dfrac{\hbar}{\epsilon_{F}}\dfrac{\partial v^{i}}{\partial\tilde{\omega}}
=N_{\hbar}\dfrac{2N_{g'}\Delta}{(\Delta^{2}+N_{\eta}^{2})^{2}}\\
&=\dfrac{2N_{\hbar}N_{g'}\Delta}{(\Delta^{2}+N_{\eta}^{2})^{2}}
=\dfrac{2N_{g}^{2}N_{\eta}\Delta}{(\Delta^{2}+N_{\eta}^{2})^{2}},\\
\end{aligned}
\end{equation}
where
\begin{equation}
\begin{aligned}
&N_{\hbar}=\dfrac{\hbar}{\epsilon_{F}},\
N_{\hbar}N_{g''}=\dfrac{\hbar^{2}g^{2}}{\epsilon_{F}^{2}}=N_{g}^{2},\\
&N_{\hbar}N_{g'}=\dfrac{\hbar^{3}g^{2}\eta}{\epsilon_{F}^{3}}=N_{g}^{2}N_{\eta},
N_{g}=\dfrac{\hbar g}{\epsilon_{F}}.
\end{aligned}
\end{equation}

The function $\chi_{0}(q,\omega)$ is a complex function consisting of both real and imaginary parts. In the long-wavelength limit $q\rightarrow 0$,
\begin{equation}
\begin{aligned}
\tilde{\chi}_{nn}^{r}(q,\omega)
=&\dfrac{nq^{2}}{m\omega^{2}}\left(1+a_{d}\dfrac{q^{2}v_{F}^{2}}{\omega^{2}}\right)\\
=&\dfrac{\hbar^{2}k_{F}^{2}n\tilde{q}^{2}}{m\epsilon_{F}^{2}\tilde{\omega}^{2}}\left(1+a_{d}\dfrac{\hbar^{2}k_{F}^{2}\tilde{q}^{2}v_{F}^{2}}{\epsilon_{F}^{2}\tilde{\omega}^{2}}\right) \\
%=\dfrac{2m\epsilon_{F}n\tilde{q}^{2}}{m\epsilon_{F}^{2}\tilde{\omega}^{2}}\left(1+a_{d}\dfrac{2m\epsilon_{F}\tilde{q}^{2}v_{F}^{2}}{\epsilon_{F}^{2}\tilde{\omega}^{2}}\right)
%=\dfrac{2n\tilde{q}^{2}}{m\epsilon_{F}\tilde{\omega}^{2}}\left(1+a_{d}\dfrac{2m\tilde{q}^{2}v_{F}^{2}}{\epsilon_{F}\tilde{\omega}^{2}}\right)
=&\dfrac{2n}{\epsilon_{F}}\dfrac{\tilde{q}^{2}}{\tilde{\omega}^{2}}\left(1+a_{d}\dfrac{4\tilde{q}^{2}}{\tilde{\omega}^{2}}\right)
=2N_{n}\dfrac{\tilde{q}^{2}}{\tilde{\omega}^{2}}\left(1+a_{d}\dfrac{4\tilde{q}^{2}}{\tilde{\omega}^{2}}\right),\\
\tilde{\chi}_{nn}^{i}(q,\omega)
=&-\dfrac{N_{\sigma}}{\sqrt{1-\left(\dfrac{q}{2k_{F}}\right)^{2}}}\dfrac{\omega}{qv_{F}} \\
=&-\dfrac{N_{\sigma}}{\sqrt{1-\left(\dfrac{\tilde{q}k_{F}}{2k_{F}}\right)^{2}}}\dfrac{\epsilon_{F}\tilde{\omega}}{\hbar\tilde{q}k_{F}v_{F}}
%=\dfrac{N_{\sigma}}{\sqrt{1-\left(\dfrac{\tilde{q}}{2}\right)^{2}}}\dfrac{\epsilon_{F}\tilde{\omega}}{\tilde{q}mv_{F}^{2}}
%=\dfrac{N_{\sigma}}{\sqrt{1-\left(\dfrac{\tilde{q}}{2}\right)^{2}}}\dfrac{\tilde{\omega}}{2\tilde{q}}
=-\dfrac{N_{\sigma}}{\sqrt{4-\tilde{q}^{2}}}\dfrac{\tilde{\omega}}{\tilde{q}}.
\end{aligned}
\end{equation}
where, $N_{n}=\dfrac{n}{\epsilon_{F}}$.

Calculate the partial derivatives of the real and imaginary parts of the response function with respect to the angular frequency,
\begin{equation}
\begin{aligned}
\dfrac{\partial\tilde{\chi}_{nn}^{r}}{\partial\omega}
=&\dfrac{\partial\tilde{\chi}_{nn}^{r}}{\partial\tilde{\omega}}\dfrac{\partial\tilde{\omega}}{\partial\omega}\\
=&N_{\hbar}\left[-\dfrac{4n}{\epsilon_{F}}\dfrac{\tilde{q}^{2}}{\tilde{\omega}^{3}}\left(1+a_{d}\dfrac{4\tilde{q}^{2}}{\tilde{\omega}^{2}}\right)
+\dfrac{2n}{\epsilon_{F}}\dfrac{\tilde{q}^{2}}{\tilde{\omega}^{2}}\left(-2a_{d}\dfrac{4\tilde{q}^{2}}{\tilde{\omega}^{3}}\right)\right]\\
&=-\dfrac{4N_{\hbar}N_{n}\tilde{q}^{2}}{\tilde{\omega}^{3}}\left(1+a_{d}\dfrac{8\tilde{q}^{2}}{\tilde{\omega}^{2}}\right), \\
\dfrac{\partial\tilde{\chi}_{nn}^{i}}{\partial\omega}
=&\dfrac{\partial\tilde{\chi}_{nn}^{i}}{\partial\tilde{\omega}}\dfrac{\partial\tilde{\omega}}{\partial\omega}
=-N_{\hbar}\dfrac{N_{\sigma}}{\sqrt{4-\tilde{q}^{2}}}\dfrac{1}{\tilde{q}}
=-\dfrac{N_{\hbar}N_{\sigma}}{\sqrt{4-\tilde{q}^{2}}}\dfrac{1}{\tilde{q}}.
\end{aligned}
\end{equation}

Expanding the effective dielectric function around the real frequency $\omega_{\text{pol}}$,
\begin{equation}
\begin{aligned}
&\epsilon_{\mathrm{eff}}(q,\omega)
=\epsilon_{\mathrm{eff}}(q,\omega_{\mathrm{pol}}-i\Gamma)  \\
%=\epsilon_{\mathrm{eff}}(q,\omega_{\mathrm{pol}})+\left.\dfrac{\partial\epsilon_{\mathrm{eff}}}{\partial\omega}\right|_{\omega_{\mathrm{pol}}}(-i\Gamma) \\
=&\epsilon_{\mathrm{eff}}^{r}(q,\omega_{\mathrm{pol}})+i\epsilon_{\mathrm{eff}}^{i}(q,\omega_{\mathrm{pol}})
+\left.\left(\dfrac{\partial\epsilon_{\mathrm{eff}}^{r}}{\partial\omega}+i\dfrac{\partial\epsilon_{\mathrm{eff}}^{i}}{\partial\omega}\right)\right|_{\omega_{\mathrm{pol}}}(-i\Gamma),
\end{aligned}
\end{equation}
The effective dielectric function is equivalent to zero means
\begin{equation}
\begin{aligned}
&\left[\epsilon_{\mathrm{eff}}^{r}(q,\omega_{\mathrm{pol}})+\left.\left(\dfrac{\partial\epsilon_{\mathrm{eff}}^{i}}{\partial\omega}\right)\right|_{\omega_{\mathrm{pol}}}\Gamma\right]  \\
&+i\left[\epsilon_{\mathrm{eff}}^{i}(q,\omega_{\mathrm{pol}})-\left.\left(\dfrac{\partial\epsilon_{\mathrm{eff}}^{r}}{\partial\omega}\right)\right|_{\omega_{\mathrm{pol}}}\Gamma\right]=0.
\end{aligned}
\end{equation}

%\subsection{}
%\subsubsection{}
% figures should be put into the text as floats.
% Use the graphics or graphicx packages (distributed with LaTeX2e)
% and the \includegraphics macro defined in those packages.
% See the LaTeX Graphics Companion by Michel Goosens, Sebastian Rahtz,
% entire page. There is no need to do explicit centering.

% \begin{figure}
% \end{figure}

% Surround figure environment with turnpage environment for landscape
% figure
% \begin{figure}
% \end{figure}

%\appendix
%\section{}

%\begin{acknowledgments}
%\end{acknowledgments}

\bibliography{ref.bib}

\end{document}